\documentclass[lettersize,journal]{IEEEtran}
\usepackage{amsmath,amsfonts}
\usepackage{algorithmic}
\usepackage{algorithm}
\usepackage{array}
\usepackage[caption=false,font=normalsize,labelfont=sf,textfont=sf]{subfig}
\usepackage{textcomp}
\usepackage{stfloats}
\usepackage{url}
\usepackage{verbatim}
\usepackage{graphicx}
\usepackage{cite}
\usepackage{makecell}
\usepackage{multirow}

\usepackage{xcolor}
\begin{document}

\title{Large Language Models for Networking: Workflow, Advances and Challenges}

\author{Chang Liu, Xiaohui Xie, Xinggong Zhang~\IEEEmembership{Senior Member,~IEEE}, Yong Cui,~\IEEEmembership{Member,~IEEE}
  \IEEEcompsocitemizethanks{
    \IEEEcompsocthanksitem C.~Liu, X.~Xie and Y.~Cui are with the Department of Computer Science and Technology, Tsinghua University, Beijing, China.
    \IEEEcompsocthanksitem X.~Zhang is with the Wangxuan Institute of Computer Technology, Peking University, Beijing, China.
    \IEEEcompsocthanksitem Yong Cui~(cuiyong@tsinghua.edu.cn) is the corresponding author.
  }
}

\maketitle

\maketitle

\begin{abstract}

The networking field is characterized by its high complexity and rapid iteration, requiring extensive expertise to accomplish network tasks, ranging from network design, configuration, diagnosis and security. 
The inherent complexity of these tasks, coupled with the ever-changing landscape of networking technologies and protocols, poses significant hurdles for traditional machine learning-based methods. 
These methods often struggle to generalize and automate complex tasks in networking, as they require extensive labeled data, domain-specific feature engineering, and frequent retraining to adapt to new scenarios.
However, the recent emergence of large language models (LLMs) has sparked a new wave of possibilities in addressing these challenges. 
LLMs have demonstrated remarkable capabilities in natural language understanding, generation, and reasoning. 
These models, trained on extensive data, can benefit the networking domain.
Some efforts have already explored the application of LLMs in the networking domain and revealed promising results.
By reviewing recent advances, we present an abstract workflow to describe the fundamental process involved in applying LLM for Networking.
We introduce the highlights of existing works by category and explain in detail how they operate at different stages of the workflow.
Furthermore, we delve into the challenges encountered, discuss potential solutions, and outline future research prospects.
We hope that this survey will provide insight for researchers and practitioners, promoting the development of this interdisciplinary research field.

\end{abstract}

\begin{IEEEkeywords}
Large Language Model, Networking, LLMN Workflow
\end{IEEEkeywords}

\section{Introduction}
\label{sec:intro}
Networks, including wide-area internet, data center networks, and satellite networks, are critical infrastructures with distinct design principles and architectures. 
Due to the complexity and rapid iteration in the field of networking, managing and maintaining these networks involves complex operations that demand rich network expertise.
Efficient solutions for network tasks such as network design, network configuration, network diagnosis, and network security are essential for ensuring network stability and scalability. 
These tasks have garnered significant attention from both academia and industry.
Although several machine learning (ML)-based methods have been proposed and shown effectiveness, they often lack generalization and necessitate repeated designs for different scenarios~\cite{boutaba2018comprehensive}.

Recently, there has been significant progress in large language models (LLMs), representing the latest advancement in generative AI.
Pretrained on extensive data, these models reveal remarkable capabilities in concept understanding, mathematical reasoning, and tool usage, enabling groundbreaking advancements in various domains, such as chip design, protein structure generation, and robot-embodied intelligence.
Inspired by these promising endeavors, some researchers have also explored the application of LLMs in the field of networking and demonstrated encouraging results.
In particular, LLMs offer significant advantages in networking due to their ability to process natural language input and output, eliminating the need for strict data modeling and feature extraction. This reduces task workloads. They can also construct logical chains for complex problem-solving by leveraging domain knowledge, and their transfer learning capability allows them to apply insights from related fields to new network challenges, enhancing their flexibility and applicability.

Given the growing interest in large language models for networking (LLMN), we believe that it is the right time to survey the current literature.
Based on the in-depth analysis of existing works, we first propose an abstract workflow to describe the fundamental process involved in LLMN.
The LLMN workflow~(Section~\ref{sec:workflow}) consists of six stages: task definition, data representation, prompt engineering, model evolution, tools integration, and validation. 
These stages aim to address complex tasks, handle diverse data types, guide LLMs in generating accurate answers, expand LLM capabilities through tool integration, and ensure performance and output validation. This workflow provides a practical roadmap for researchers entering the field of LLMN.

We then conduct a targeted review (Section~\ref{sec:advances}) of notable advancements in networking facilitated by LLMs, predominantly sourced from recent publications.
For example, LLMs can assist in designing algorithms and managing network topologies~\cite{he2024llm,mani2023enhancing}, enhance configuration efficiency while minimizing errors~\cite{mondal2023llms,lian2023configuration}, identify hidden patterns or anomalies~\cite{zhou2023towards,kotaru2023adapting}, and introduce novel strategies for improving network security~\cite{meng2024large}.
We categorize these advancements into several key fields of networking, including network design, network configuration, network diagnosis, and network security, and elaborate on how they are performed at each stage of the LLMN workflow.

In addition to the literature review, we also identify several challenges~(Section~\ref{sec:challenges}) that necessitate further research.
These challenges span across different stages of the LLMN workflow and can be categorized into six aspects: intelligent planning, multimodal data understanding, network-specific LLM construction, autonomous tool utilization by LLMs, reliability and safety assurance, and efficiency and real-time performance improvement. 
Potential approaches to address them are also discussed.

Compared to a prior publication that merely focuses on a domain-adapted LLM framework with access to various external network tools~\cite{huang2023large}, our comprehensive survey presents a condensed overview of existing LLMN research and the lessons learned from it.
Moreover, based on the understanding of existing research, we have abstracted a universal workflow that includes essential mechanisms such as validation and feedback.
This survey aims to provide a foundational and practical research roadmap for newcomers to the LLMN field, while also enabling experienced network professionals to quickly grasp the transformative potential of LLMs within their domain. Additionally, we hope this article serves as a catalyst for further research endeavors by inspiring networking and artificial intelligence (AI) experts to delve into more profound investigations.

\section{Basic Workflow for LLMN}
\label{sec:workflow}

\begin{figure*}[!t]
    \centering
    \includegraphics[width=1.0\textwidth]{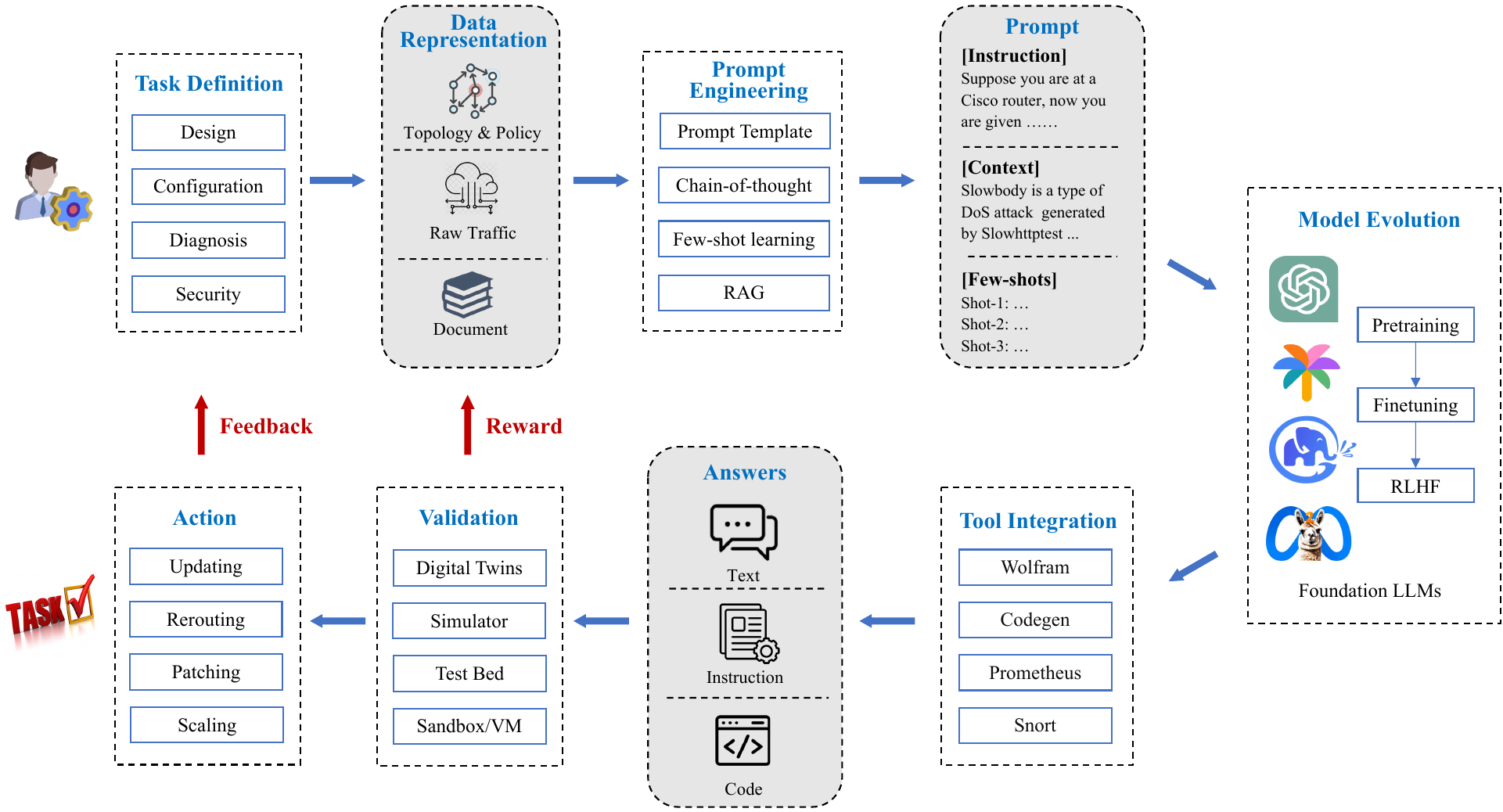}
    \caption{The typical workflow of large language models for networking.}
    \label{fig_1}
\end{figure*}

In this section, we present an abstract workflow for integrating LLMs into networking tasks. The proposed workflow, illustrated in Figure~\ref{fig_1}, consists of six key stages: task definition, data representation, prompt engineering, model evolution, tool integration, and validation. While these stages are interconnected, they are not fully coupled, allowing for flexibility and adaptability in the implementation process. Each stage encompasses a range of unique techniques and considerations, which we will explore in depth.

\textbf{Task Definition.} LLMs are designed to process natural language, so it's important to present intricate networking tasks in a way that LLMs can understand. One approach is to assign personas to LLMs and provide initial global instructions that clarify the task's goals and requirements~\cite{zhou2023towards, hamadanian2023holistic}. 
While large language models can capture contextual information to some extent, their short-term memory is limited. As a result, they may not perform well when dealing with longer text sequences or tasks that require long-term planning. 
To overcome this limitation, decomposing complex tasks into sub-tasks can be helpful. This allows for detailed inputs and expected outputs for each sub-task.
For example, the Incident Management (IM) process is time-consuming and labor-intensive due to the possible root causes being varied. Instead of directly incorporating LLMs to solve the end-to-end problem, the IM process can be broken down into three sub-tasks: hypothesis formation, hypothesis testing, and mitigation planning\cite{hamadanian2023holistic}.

\textbf{Data Representation.} 
The objective of this stage is to handle heterogeneous data, including graph-based network topology, control plane policies, binary traffic data from the data plane, and text-based domain knowledge. 
Translating domain-specific information into natural language is an intuitive approach, e.g., describing the characteristics of traffic using natural language, but it may lack generality and lead to information loss. 
Conversely, XML, YAML, and JSON formats offer high flexibility in representing complex data structures and are commonly used for data transmission and storage in various applications and systems. 
For textual documents, a suitable approach is to store the preprocessed information in vector databases (e.g. Pinecone~\footnote{Pinecone: https://www.pinecone.io/}, Milvus~\footnote{Milvus: https://milvus.io/}), which might enable efficient similarity searches and allow for effective retrieval of relevant knowledge. For example, to facilitate retrieving and analyzing operator data, comprehensive textual descriptions of various metrics and custom functions can be stored in vector databases for semantic search~\cite{kotaru2023adapting}.

\textbf{Prompt Engineering.} 
Crafting effective prompts is crucial for guiding LLMs to produce desired outputs in networking tasks. Prompts should include descriptions of the network environment, such as network topology, device configurations, and protocol specifications. 
This helps LLMs understand the task context and generate outputs that align with the network environment. However, due to prompt length limitations, we need to include only closely relevant information, which can be achieved through Retrieval Augmented Generation (RAG) techniques.
Additionally, LLM outputs often serve as inputs for other tools or instructions for network device operations, requiring adherence to specific requirements. To address these challenges, few-shot learning and Chain-of-thought are proposed to guide LLMs in generalizing and generating accurate results with limited input-output examples, proving to be highly effective. 
For example, to generate machine-readable grammars for different message types in the RTSP protocols, the prompt context includes the PLAY client request grammar for RTSP and the GET client request grammar for HTTP~\cite{meng2024large}.

\textbf{Model Evolution.}
Foundation LLMs play a crucial role in the LLMN workflow, and their capabilities significantly impact task completion. However, these models, pre-trained on general data, have limitations in addressing networking domain tasks. 
Enhancing LLM performance can be achieved by incorporating additional network-specific data or knowledge during various stages of training. Fine-tuning LLMs using the latest network data is a common practice, considering the high cost of training from scratch and the dynamic nature of network environments. 
This approach enables adaptation to network dynamics, providing relevant and effective solutions to current network challenges~\cite{wu2024large}. Moreover, techniques such as Human Feedback Reinforcement Learning (RLHB) allow models to learn from the expertise of network professionals, generating outputs that align better with the context and values of the networking domain.

\textbf{Tool Integration.} LLMs excel at reading comprehension and semantic analysis, generating answers based on a pre-established probability distribution in an auto-regressive manner. However, they may struggle with symbolic reasoning and mathematical deduction. To enhance the accuracy of numerical results, crucial for decision-making in networking, one approach is to leverage mathematical tools like Wolfram~\footnote{Wolfram: https://www.wolframalpha.com/} or utilize Python code for data analysis and calculation~\cite{kotaru2023adapting, mani2023enhancing}. 
In addition to general tools, detailed descriptions and usages of domain-specific tools can be provided to enable LLMs to automatically determine and select the appropriate tool based on different task requirements. For example, when it comes to synthesizing router configurations, LLMs alone are not effective. However, by incorporating existing configuration verifiers like Batfish~\footnote{Batfish: https://batfish.org/} and Campion~\footnote{Campion: https://github.com/atang42/batfish/tree/rm-localize}, which offer actionable feedback to address uncertainties and invalidities, LLMs can significantly reduce the need for human intervention~\cite{mondal2023llms}.

\textbf{Validation.} Validation is an indispensable stage in the LLMN workflow to assess the correctness and security of LLM-generated outputs, which can include textual analysis reports, graph manipulation code, and network device instructions. 
To minimize syntax and semantic errors in LLM-generated code, techniques like self-consistency and self-debugging can be employed~\cite{mani2023enhancing}. 
However, code that is syntactically and semantically correct may still lack proper functionality. Simulation and emulation are commonly used techniques that can help verify whether the code and operations generated by LLM have achieved the expected functionality.
Furthermore, before deploying LLM-generated code and operations in a production environment, thorough risk monitoring and control measures should be implemented.
An execution sandbox, for instance, creates an isolated environment to run uncertain or potentially harmful code, mitigating any adverse effects on the actual network~\cite{mani2023enhancing}. 

The key stages mentioned above collectively determine the efficiency and performance of leveraging LLMs in handling various network tasks.
Task definition focuses on describing and decomposing complex tasks, while data representation handles different data types.
Prompt engineering guides LLMs in understanding tasks, and model evolution concentrates on the model's performance.
Tool integration extends LLM's capabilities, and validation is used to assess the correctness and security of the output.
Furthermore, we underscore that iterative optimization can occur within each of these stages through two distinct feedback mechanisms: an internal loop driven by validation outcomes and an external loop guided by task execution results.
Validation outcomes can steer the continuous learning and enhancement of LLMs, leading to the generation of outputs that are more precise and high-performing.
Task execution feedback, on the other hand, can assist network administrators in refining execution procedures tailored to various network task scenarios, thereby promoting smoother integration of LLMs.

\section{Overview of Recent Advances}
\label{sec:advances}

The networking domain is a highly complex and multifaceted field that requires a significant amount of specialized knowledge and skill. 
The notable success and widespread adoption of LLMs have greatly attracted the attention of the networking community, leading to a variety of attempts to reform the existing working paradigms in different subfields of networking, including network design, configuration, diagnosis, and security.
In this section, we provide a general overview of the up-to-date advances in applying LLMs to the networking domain. 
Table~\ref{tab:advancesandLLMN} outlines the approaches taken by existing studies at every stage of the LLMN workflow.

\begin{table*}[!t]
    \centering
    \caption{Relationships between the latest advances and key stages in LLMN workflow.}
    \begin{tabular}{|m{1.5cm}<{\centering}|m{1.6cm}<{\centering}|m{2cm}<{\centering}|m{2.4cm}<{\centering}|m{2cm}<{\centering}|m{2cm}<{\centering}|m{1.5cm}<{\centering}|m{1.6cm}<{\centering}|}
    \hline
    \thead{Research\\Work}&\thead{Application\\Scenario}&\thead{Task\\Definition}&\thead{Data\\Representation}&\thead{Prompt\\Engineering}&\thead{Model\\Evolution}&\thead{Tools\\Integration}&Validation\\
    \hline
    He et al.~\cite{he2024llm}&\multirow{2}{*}[-5ex]{Design}&Designing Adaptive Bitrate Algorithms&Rewrite Pensieve’s original code of state creation and network architectures into a function respectively&CoT \& Prompt Template&Off-the-shelf GPT-3.5 \& GPT-4& / & Evaluation of various candidate algorithms in a network simulator\\
    \cline{1-1} \cline{3-8}
    Mani et al.~\cite{mani2023enhancing}&&Generating code for graph analysis and manipulation&Representing network topology in different formats&Application Prompt Generator \& Code-Gen Prompt Generator&Off-the-shelf LLMs like GPT-4, GPT-3, text-davinci-003 and Google Bard&NetworkX with flexible API&Execution in a sandbox environment\\
    \hline
    Mondal et al.~\cite{mondal2023llms}&\multirow{2}{*}[-6ex]{Configuration}&Synthesizing router configurations&Representing network topology and connection with JSON&One shot learning \& Prompt Template&Off-the-shelf GPT-4&Syntax verifier \& Semantics verifier&Manual correction\\
    \cline{1-1} \cline{3-8}
    Lian et al.~\cite{lian2023configuration}&&Configuration validation to detect errors&Maintaining a database of labeled valid configurations and misconfigurations&Prompt template, Few-shot learning \& RAG&Off-the-shelf LLMs like Code-Davinci-002, GPT-3.5-turbo, GPT-4 and Babbage-002&/&Aggregation of outputs from various LLMs by voting\\
    \hline
    Kotaru~\cite{kotaru2023adapting}&\multirow{2}{*}[-7ex]{Diagnosis}&Retrieval and analysis of operator data&Building a domain-specific database to store definitions of various metrics and custom functions&Few-shot learning \& RAG&Off-the-shelf LLMs like GPT-4, GPT3.5-turbo and text-curie-001&/&Execution in a sandbox environment \& Creating GitHub issues to attain feedback from experts\\
    \cline{1-1} \cline{3-8}
    Zhou et al.\cite{zhou2023towards}&&Assisting in conducting network incident investigations as a research assistant&Saving the most relevant domain knowledge in a JSON file&CoT \& RAG&Off-the-shelf GPT-4&Search engines&Self-test \& Self-learning\\
    \hline
    Meng et at.~\cite{meng2024large}&\multirow{2}{*}[-6ex]{Security}&Assisting in protocol fuzzing&Representing protocol formats with specific strings patterns&Few-shot learning&Off-the-shelf GPT-3.5-turbo&/&Self-consistency check\\
    \cline{1-1} \cline{3-8}
    Wang et al.~\cite{wang2024shieldgpt}&&Providing in-depth analysis and customized mitigation recommendations for DDoS attack&Converting raw traffic into LLM-understandable texts&Prompt template&Off-the-shelf GPT-4&/&/\\
    \hline
    \end{tabular}
    \label{tab:advancesandLLMN}
\end{table*}

\subsection{Network Design}
The Network Design task involves detailed planning processes like protocol selection, bandwidth allocation, and network topology optimization. The challenge lies in making optimal decisions amidst ever-evolving technology, complex requirements, and limited resources.
Traditionally, network engineers have depended on their expertise, manual fine-tuning, and basic tools for network design and optimization. However, LLMs offer a promising avenue by combining extensive networking knowledge with advanced reasoning and generation capabilities. 
LLMs can enhance network design by proposing tailored strategies for specific network goals and facilitating the automatic evaluation of various design options.

As a part of initial investigations, several pieces of literature focus on the critical importance of data transmission and topology management in network architecture. 
He et al.~\cite{he2024llm} proposed leveraging LLM to develop ABR (Adaptive Bitrate) algorithms, a pivotal component for dynamically adjusting video quality in streaming media delivery. Given the challenge of directly crafting high-quality algorithms with LLM, they incorporated the classic Pensieve algorithm as part of the input prompts. Their objective was to ask LLM to enhance the original algorithm and generate a range of candidate algorithms. As not all these algorithms exhibited significant enhancements, evaluating these algorithms in a network simulator was imperative to identify the most effective one. Experimental findings demonstrate LLM's capability to produce outstanding algorithms tailored to diverse network conditions.
Mani et al.~\cite{mani2023enhancing} have investigated the use of LLMs for code generation in graph analysis and manipulation for topology management. They proposed a structured framework consisting of application wrappers, application prompt generators, code-generation prompt generators, and execution sandboxes. This framework generates user query-based prompts for specific network applications, guides LLMs in producing practical code, and provides a secure execution environment for this code. Their findings reveal that this method significantly improves code quality over direct LLM processing of raw network data, although code quality tends to decline with increasing task complexity. The study suggests that future efforts should aim at enhancing LLMs' capabilities in managing complex network tasks.

\subsection{Network Configuration}

The evolution of networking and cloud computing technologies has led to the proliferation of network devices, each running various applications to deliver services. Ensuring these devices operate correctly is critically dependent on precise network configuration, a task complicated by the myriad of settings and parameters that demand a deep grasp of networking concepts and operational details. Moreover, the dynamic nature of network conditions and business objectives requires configurations to be regularly updated to align with emerging needs.
Traditionally, network configurations have been manually crafted and verified, a process that is not only time-consuming but also prone to errors, making it ill-suited for the dynamic and complex nature of modern networks and their rapidly evolving requirements.

Inspired by the advanced capabilities of LLMs in aiding with programming tasks, Mondal et al.~\cite{mondal2023llms} have explored the application of LLMs, such as GPT-4, in synthesizing network router configurations.
They introduced a novel approach, \textit{Verified Prompt Programming}, which combines LLMs with a verification system to automatically correct errors using localized feedback, thus enhancing the accuracy of the generated configurations. 
Their research focused on practical applications, such as converting router configurations from Cisco to Juniper format and implementing a no-relay policy across routers. 
Despite the advancements, they emphasize the continued importance of human oversight in refining LLM-generated configurations to ensure the precision required for router setups.
Aimed at enhancing the process of configuration validation prior to deployment., Lian et al.~\cite{lian2023configuration} have developed a novel LLM-based framework, named Ciri.
This approach marks a significant shift from traditional methods that rely heavily on manual checks or extensive engineering, towards a more efficient and automated solution. 
Ciri leverages advanced prompt engineering with few-shot learning, drawing on existing configuration data to identify and explain potential misconfigurations from either complete files or diffs.
Furthermore, Ciri enhances reliability by aggregating insights from multiple LLMs, such as Code-Davinci-002, GPT-3.5-turbo, and GPT-4, through a voting mechanism. This strategy effectively mitigates issues like hallucination and inconsistency, common in LLM outputs. However, while Ciri excels at detecting syntax and range errors, it encounters challenges with configurations that involve intricate dependencies or specific software version requirements. Overcoming these limitations may involve retraining or fine-tuning LLMs with targeted data on dependencies and versioning.

\subsection{Network Diagnosis}
Networks are pivotal in driving social and business advancements, with even minor dysfunctions potentially causing substantial losses. Fault diagnosis is critical for addressing network issues, encompassing data collection (like system logs and traffic data), analysis (to spot abnormal patterns or faults), identifying root causes, and implementing fixes. Given the complexity of network environments and the variety of potential issues, fault diagnosis presents significant challenges. Typically, administrators might sift through extensive monitoring data and consider numerous potential causes, a process that demands a deep understanding of diverse systems. However, LLMs offer a promising solution, capable of processing vast datasets and identifying hidden patterns or anomalies, thus potentially enhancing the efficiency and precision of fault diagnosis processes.

In commercial network operations, managing thousands of counters and metrics for data analysis is a complex task. To reduce reliance on specialists and expedite issue resolution, Kotaru~\cite{kotaru2023adapting} introduced the Data Intelligence for Operators Copilot (DIO Copilot), a natural language interface utilizing LLMs for efficient data retrieval and analysis. DIO Copilot enhances metric extraction through semantic search and employs few-shot learning, enabling LLMs to accurately respond to user queries with minimal examples. Notably, it features a mechanism for generating GitHub issues to foster expert collaboration and knowledge updates. When evaluated against leading natural language interfaces for databases, DIO Copilot demonstrated superior performance, achieving 66\% execution accuracy on a benchmark dataset. Besides data retrieval and analysis, recent studies~\cite{hamadanian2023holistic, zhou2023towards} have explored the use of LLMs in incident management and investigation. 
Hamadanian et al.~\cite{hamadanian2023holistic} have proposed a holistic theoretical framework aimed at building an AI helper for incident management. 
This framework is guided by three fundamental principles: iterative prediction, reliable \& safe, and adaptive. 
In contrast to the one-shot feed-forward model, iterative prediction can successfully handle more complex (often more impacting) or novel incidents by hypothesizing, testing, and re-evaluating of its decisions in a feedback loop.
Concurrently, Zhou et al.~\cite{zhou2023towards} developed an interactive software agent aimed at aiding in the investigation of internet incidents. 
This agent is structured around four key components: role definition, information retrieval, knowledge memory, and knowledge testing with self-learning capabilities. Utilizing AutoGPT~\footnote{AutoGPT: https://github.com/Significant-Gravitas/AutoGPT}, the agent autonomously gathers relevant information from the internet, enhancing its reasoning and query response abilities. It continuously enriches its knowledge base until it achieves sufficient confidence to function akin to a researcher. 
In a practical test, an agent named Bob was trained to autonomously learn about solar storm events, successfully providing expert-level insights into their effects on the internet, and demonstrating capabilities comparable to professional researchers.

\subsection{Network Security}
Network security is a critical concern for individuals, companies, and governments, involving a continuous process of offense and defense. Defenders employ various measures to strengthen defenses, such as conducting regular risk assessments and vulnerability scans, deploying advanced security technologies, and implementing strict security policies. Attackers, on the other hand, continuously develop advanced attack methods, leveraging automated tools and launching large-scale botnet-driven distributed denial-of-service (DDoS) attacks. The emergence of LLMs introduces new approaches for both offense and defense, marking a new stage in network security.

Mutation-based protocol fuzzing is a security testing technique that assesses protocol implementations by fuzzing seed messages, aiming to uncover potential vulnerabilities. However, the limited diversity of seed messages poses a constraint. To address this limitation, Meng et al.~\cite{meng2024large} proposed ChatAFL, a novel engine that enhances existing protocol fuzzing tools using LLMs. 
ChatAFL builds upon AFLNet and introduces machine-readable protocol grammar extraction to enrich seed inputs' diversity. Moreover, when the fuzzer exhausts exploration of new states, ChatAFL interacts with the LLM to generate client requests that induce server transitions. Experimental results demonstrate that compared to the baseline AFLNet, ChatAFL achieves improved state transition coverage, state coverage, and code coverage, while also identifying more security vulnerabilities.
Considering existing AI-driven methods for DDoS defense have certain limitations in practical application, including lack of explainability and lack of mitigation instructions, Wang et al.~\cite{wang2024shieldgpt} have proposed a LLM-based DDoS defense framework named ShieldGPT to solve these shortcomings. 
ShieldGPT comprises four modules: attack detection, traffic representation, domain-knowledge injection, and role representation. It transforms raw traffic data into LLM-friendly text, maintaining essential features. By integrating this data with domain knowledge using prompt templates, ShieldGPT enables the LLM to elucidate attack behaviors and propose specific mitigation actions. 
Preliminary experiments demonstrate that ShieldGP can provide in-depth attack analysis and customized mitigation recommendations, laying the foundation for establishing a truly autonomous DDoS defense system.

\section{Challenges and Future Prospects}
\label{sec:challenges}
The previous research carried out in various aspects aligns well with the proposed LLMN workflow, indicating its universal applicability and providing guidance for network researchers to advance their studies. However, upon analyzing these existing works, we identify several pressing challenges within the aforementioned workflow that require further research efforts. In this section, we present the following challenges and prospects.

\subsection{Intelligent Planning}
Most of the current research focuses on relatively straightforward tasks, while complex tasks that involve long-term objectives and multi-step decision-making still require manual intervention and lack an end-to-end solution. 
Intelligent planning capabilities are essential for LLMs to effectively generate plans and take appropriate actions to accomplish these complex tasks. 
However, research on enhancing the performance of LLMs in tackling such complex network tasks is currently limited. 
Future work should concentrate on formalizing the task execution process in the networking domain and establishing a comprehensive task library~\cite{paranjape2023art}.
This task library would enable researchers to utilize task rewards to improve the LLM's planning and problem-solving abilities, establishing a self-improving feedback loop~\cite{ge2024openagi}.

\subsection{Understanding Multimodal Data}
A prevalent modal mismatch exists between text-based natural language and heterogeneous network data. 
Previous studies have demonstrated that manual processing or script construction is required to convert multimodal network data into text, in a way LLMs can understand.
This approach incurs high costs and lacks adaptability to task variations.
To address this, researchers are exploring lightweight preprocessing and representation learning techniques to map multimodal data into a shared vector space or leverage linear projection to map features extracted by different encoders into LLM's token space~\cite{wu2024large, zhang2023robust}.
These efforts serve as a promising starting point. 

\subsection{Network-specific LLMs}
Existing approaches are to guide LLMs originally designed for general domains to better perform network-related tasks through prompt engineering.
However, a more effective strategy would be to construct LLMs specialized for the network domain, introducing heterogeneous data sources in the pretraining stage and establishing strong semantic connections between them.
Based on such a powerful LLM, we can fine-tune it for various targeted tasks to achieve better results.
Additionally, given the structural disparities between network and text data, as well as the spatiotemporal constraints in network environments, transformer-based models may not be the most suitable architecture for these scenarios. Notably, new frameworks like Mamba have been proposed and deserve attention in the realm of LLMN.

\subsection{Autonomous Tools Utilization}
In the field of networking, numerous valuable tools enhance efficiency and accuracy in handling specific tasks.
However, LLMs currently face limitations in autonomously leveraging these tools. To address this, it is essential to establish a well-structured tool library that showcases detailed use cases of diverse tools and design a unified interface between LLMs and the tools. The interface should fulfill the following requirements:
(1) Standardized input and output formats, enabling LLMs to seamlessly invoke different tools and synthesize the execution results;
(2) Flexible scalability to facilitate the integration of new tools and features, ensuring adaptability to evolving demands and technologies.

\subsection{Reliability and Safety}
Ensuring the reliability and security of LLM applications is a critical challenge that needs to be addressed. 
While there is existing research focusing on the accuracy and consistency of LLM outputs, there is a relative scarcity of studies on controlling and mitigating risks associated with implementing LLM outputs in real network operations. We propose that integrating LLM with validation environments such as digital twins can enhance the reliability and security of real network environments. These validation environments can simulate system behavior, operate synchronously with real systems, monitor system status, and provide real-time feedback. This integration can effectively meet the requirements for practical LLM applications in the network domain.

\subsection{Efficiency and Real-time Performance}
Many network tasks such as resource scheduling and fault diagnosis have time constraints, while LLMs have relatively slow inference speeds. Additionally, ensuring the reliability and safety of LLM outputs often requires manual verification and iterative execution, posing challenges for meeting real-time requirements. To enhance the efficiency and real-time performance of LLMs in executing network tasks, future research can focus on two perspectives. Firstly, implementing techniques like model compression and optimization to reduce computational load and accelerate inference speed. Secondly, designing automated task execution and validation processes to minimize human intervention. 

\section{Conclusion}
\label{sec:conclusion}
In this paper, we provide a foundational workflow as a practical guide for researchers to explore novel applications of LLMs in networking research. For a deeper comprehension, we summarize the recent advances of LLMs in the networking domain, covering various important subfields including design, configuration, diagnostics, and security. Furthermore, there are still many unresolved challenges and we shed light on directions for further research efforts aimed at better integrating LLMs with networking.


\bibliographystyle{IEEEtran}

\bibliography{IEEEabrv, ref}

\begin{IEEEbiographynophoto}{Chang Liu} is currently pursuing a Ph.D. degree with the Department of Computer Science and Technology, Tsinghua University, Beijing, China. He received the B.E. and M.E. degrees in the School of Computer Science, National University of Defense Technology, Changsha, China, in 2013 and 2015, respectively. His current research interests include network security and artificial intelligence.
\end{IEEEbiographynophoto}

\begin{IEEEbiographynophoto}{Xiaohui Xie} received the B.E. and Ph.D. degrees in computer science and engineering from Tsinghua University, China, in 2016 and 2021, respectively. He is currently an Assistant Professor at the Computer Science Department, Tsinghua University.
His research interests include Network Anomaly Detection, Information Retrieval and Artificial intelligence.
\end{IEEEbiographynophoto}

\begin{IEEEbiographynophoto}{Xinggong Zhang (Senior Member, IEEE)} received the Ph.D. degree from the Department of Computer Science, Peking University, Beijing, China, in 2011. He is currently an Associate Professor with the Wangxuan Institute of Computer Technology, Peking University. Before that, he was a Senior Researcher with the Founder Research and Development Center, Peking University, from 1998 to 2007. He was a Visiting Scholar with the Polytechnic Institute of New York University from 2010 to 2011. His research interests include the modeling and optimization of multimedia networks, VR/AR/video streaming, and satellite networks.
\end{IEEEbiographynophoto}

\begin{IEEEbiographynophoto}{Yong Cui (Member, IEEE)} received the B.E. and Ph.D. degrees in computer science and engineering from Tsinghua University, China, in 1999 and 2004, respectively. He is currently a Full Professor with the Computer Science Department, Tsinghua University. His major research interests include mobile cloud computing and network architecture. He served or serves on the editorial boards of the IEEE TPDS, IEEE TCC, IEEE Internet Computing, and the IEEE Network.
\end{IEEEbiographynophoto}

\vfill

\end{document}